\documentclass[aps,pra,reprint,floatfix]{revtex4-1}
\usepackage{graphicx}
\usepackage{amsmath}
\usepackage{amssymb}
\usepackage{mathrsfs}
\usepackage{braket}
\usepackage{natbib}
\newcommand{\st}[1]{}
\newcommand{\textcolor}[2]{#2}
\begin{document}
\begin{titlepage}
\title{Quantum Time Crystal By Decoherence:\\ Proposal With Incommensurate Charge Density Wave Ring}
\author{K. Nakatsugawa$^1$, T. Fujii$^3$, S. Tanda$^{1,2}$}
\vspace*{.2in}
\affiliation{
$^1$Department of Applied Physics, Hokkaido University, Sapporo 060-8628, Japan\\
$^2$Center of Education and Research for Topological Science and Technology, Hokkaido University, Sapporo 060-8628, Japan\\
$^3$Department of Physics, Asahikawa Medical University, Asahikawa 078-8510, Japan
}
\date{\today}
\vspace*{.3in}
\begin{abstract}
We show that time translation symmetry of a ring system with a macroscopic quantum ground state is broken by decoherence. In particular, we consider a ring-shaped incommensurate charge density wave (ICDW ring) threaded by a fluctuating magnetic flux: the Caldeira-Leggett model is used to model the fluctuating flux as a bath of harmonic oscillators. We show that the charge density expectation value of a quantized ICDW ring coupled to its environment oscillates periodically. The Hamiltonians considered in this model are time independent unlike ``Floquet time crystals" considered recently. Our model forms a metastable quantum time crystal with a finite length in space and in time.
\end{abstract}
\nopagebreak
\maketitle
\end{titlepage}
\section{INTRODUCTION}
\begin{figure}
\centering
\includegraphics[width=\linewidth]{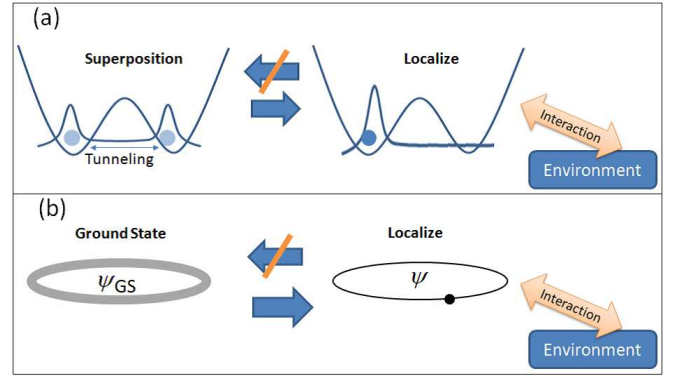}
\caption{The concept of translation symmetry breaking by decoherence is illustrated. (a) A simplified version of the two-state system considered by Leggett \textit{et al.}\cite{LeggettTwoState}. A particle can tunnel through a potential barrier and exist at two states simultaneously. However, if this system starts to interact with its surrounding environment, then the particle will localize at one of the states. (b) Similarly, the ground state of a free particle confined on a ring is a plane wave state. Coupling to environment will localize the particle and break rotational symmetry. This ``particle" corresponds to the phase of an incommensurate charge density wave ring (ICDW ring) in our model (FIG. \ref{FluctuatingB}).}
\label{Symm.Break by Decoh.}
\end{figure}

\begin{figure}
\centering
\includegraphics[width=\linewidth]{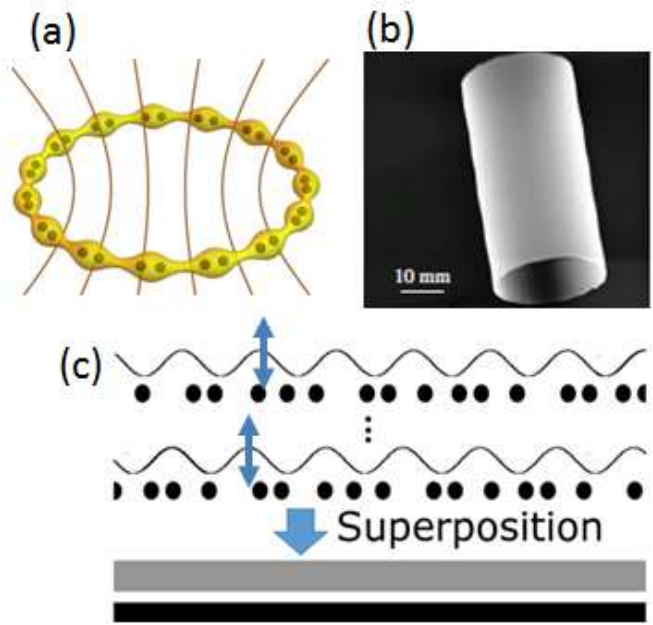}
\caption{(a) We consider an incommensurate CDW ring threaded by a fluctuating magnetic flux. This figure shows a commensurate CDW with $\lambda/a=2$ because it is easier to visualise. The wave (typically $\sim 10^5$ wavelengths) represents the charge density and the dots represent the atoms of a quasi-one dimensional crystal.(b) ICDW ring crystals {such as monoclinic TaS$_3$ ring crystals and NbSe$_3$ ring crystals }can be produced experimentally \cite{ring,Matsuura,ABCDW}. Our model can be tested provided clean ring crystals with almost no defects and impurities can be produced. (c)The ground state of an isolated quantized ICDW ring is a superposition of periodically oscillating ICDWs, hence the oscillation is unobservable. Coupling to environment (fluctuating magnetic flux) will break the superposition and the oscillation becomes apparent.}
\label{FluctuatingB}
\end{figure}

\noindent 
The original proposal of a quantum time crystal (QTC) given by Wilczek \cite{QTC} and Li \textit{et al.} \cite{Wigner} is a quantum mechanical ground state which breaks time translation symmetry. In this QTC ground state there exists an operator $\hat Q$ whose expectation value oscillates permanently with a well-defined ``lattice constant" $P$, that is, with a well-defined period. 
\\
\indent
Volovik \cite{Volovic} relaxed the condition of permanent oscillation and proposed the possibility of effective QTC, that is, a periodic oscillation in a metastable state such that the oscillation will persist for a finite duration $\tau_Q\gg P$ in the time domain and will eventually decay.
\\
\indent
In this paper we promote Volovik’s line and consider the possibility of a metastable QTC state without spontaneous symmetry breaking: We consider symmetry breaking by \textit{decoherence} of a macroscopic quantum ground state (FIG. \ref{Symm.Break by Decoh.}). Decoherence is defined as the loss of quantum coherence of a system coupled to its environment. Coupling to environment will inevitably introduce friction to the system such that the oscillation will eventually decay at $t=\tau_\text{damp}$. However, for $t<\tau_Q\ll\tau_\text{damp}$ the oscillation period $P$ is well defined. If friction is sufficiently weak such that $P\ll\tau_Q$, then we have a model of effective QTC with life time $\tau_Q$. 
\\
\indent
Our model consists of a ring-shaped incommensurate charge density wave (ICDW ring) threaded by a fluctuating magnetic flux (FIG. \ref{FluctuatingB}(a)). A charge density wave (CDW) is a periodic (spatial) modulation of electric charge density which occurs in quasi-one-dimensional {crystals} \cite{Gruner2,*Sambongi,*Monceau}: the periodic modulation of the electric charge density occurs due to electron-phonon interaction. \textcolor{red}{If the ratio between the CDW wavelength $\lambda$ and the lattice constant $a$ of the crystal is a simple fraction like $2$, $5/2$ etc, then the CDW is commensurate with the underlying lattice. A commensurate CDW cannot move freely because of commensurability pinning, i.e. the CDW phase is pinned by ions’ position in the crystal. On the other hand, if $\lambda/a$}\st{If the ratio between the CDW wavelength $\lambda$ and the lattice constant $a$ of the crystal} is effectively an irrational number \cite{Sambongi}, then the CDW is incommensurate with the underlying lattice. An ICDW ring with a radius of $10\mu$m, for instance, contains approximately $10^5$ wavelengths \cite{Zettl}, so it is possible that $a/\lambda$ is very close to an irrational number. \textcolor{red}{See the discussion section for an elaboration of this assumption.}
The sliding of an ICDW {without pinning} is described by a gapless Nambu-Goldstone (phason) mode \cite{Gruner2} and the energy of an ICDW is independent of its phase (i.e. position), which implies that the expected ground state of an ICDW ring is a superposition of ICDWs with different phases. 
Ring-shaped crystals and ring-shaped (I)CDWs have been produced \cite{ring} (FIG. \ref{FluctuatingB} (b)). The presence of circulating CDW current \cite{Matsuura} and Aharonov-Bohm oscillation (evidence of macroscopic wave function) \cite{ABCDW} are verified experimentally. 
\\
\indent 
We show in section \ref{Without_Dissipation} that the charge density expectation value of an isolated ICDW ring {with moment of inertia $I$ is periodic in time with period $P=4\pi I/\hbar$}. This periodicity is a consequence of the uncertainty relation on $S^1$ (ring). However, this oscillation becomes unobservable at ground state because the ground state of an isolated ICDW ring is a plane wave state, \textit{i.e.} a coherent superposition of ICDWs with different phases. Therefore, in section \ref{With_Dissipation} we use the Caldeira-Leggett model \cite{Caldeira,Weiss} to show that time translation symmetry is broken by decoherence. More precisely, the {superposition is broken by decoherence} and the amplitude of the ICDW oscillates periodically (FIG. \ref{FluctuatingB} (c)). If the ICDW ring weakly couples to its environment then this state is a metastable ground state. Therefore, our model forms an effective QTC with a finite length in space and in time.
\\
\indent
Before developing our main arguments, we compare our work to recent developments of QTC. In analogy with spatial crystal, the original proposal of QTC is based on the spontaneous breaking of time translation symmetry. However, Bruno \cite{Bruno3} and Watanabe and Oshikawa \cite{Watanabe} theoretically proved that spontaneous breaking of time translation symmetry cannot occur at ground state. 
Recently, it was shown that there is a notion of spontaneous breaking of time translation symmetry in periodically driven (Floquet) states \cite{FTCTheory,*DTCYao,*Prethermal} and this idea was proved experimentally\cite{FTCChoi,*FTCZhang}. On the other hand, the periodic oscillation we consider in this paper is inherent to ring systems with a macroscopic wave function.
\section{Ground State of an Isolated ICDW Ring}\label{Without_Dissipation}
\subsection{Classical Theory of ICDW Ring}
It is well known that the electric charge density of a quasi-one-dimensional crystal becomes periodic by opening a gap at the {Fermi wave number $k_\mathrm{F}$} and form a charge density wave (CDW) {ground} state with a wavelength $\lambda=\pi/k_\mathrm{F}$ \cite{Gruner2}. Consider a CDW formed on a ring-shaped quasi-one-dimensional crystal with radius $R$. 
The order parameter of this CDW ring is a complex scalar $\Delta(x,t)=|\Delta(x,t)|\exp[i\theta(x,t)]$, where $|\Delta(x,t)|$ is the size of the energy gap at $\pm k_\mathrm{F}$, $\theta(x,t)$ is the phase of the CDW, $x\in[0,2\pi R)$ is the coordinate on the crystal, and $t$ is the time coordinate. The charge density is given by
\begin{equation}
n(x,t)=n_0+n_1\cos[2k_\text Fx+\theta(x,t)]
\label{chargedensity}
\end{equation}
where $n_0$ is the average charge density and $n_1$ is the amplitude of the wave. Bogachek \emph{et al.} \cite{Bogachek1} derived the following Lagrangian density of the phase of a ring-shaped incommensurate CDW (ICDW ring) threaded by a magnetic flux
\begin{equation*}
{\mathscr{L}_0\left(\frac{\partial\theta}{\partial t},\frac{\partial\theta}{\partial x}\right)}=\frac{N_0}{2}\left[\left(\frac{\partial \theta}{\partial t}\right)^2-c_0^2\left(\frac{\partial \theta}{\partial x}\right)^2\right]+\frac{eA}{\pi}\frac{\partial \theta}{\partial t}
\end{equation*}
where $A$ is the magnetic vector potential, 
$N_0=v_\text F^2\hbar^2N(\varepsilon_\mathrm{F})/(2c_0^2)$, $N(\varepsilon_\mathrm{F})$ is the density of states of electrons at the Fermi level per unit length and per spin direction, $v_\text F$ is the Fermi velocity of the crystal, $c_0
=\sqrt{m/m^\ast}v_\text F$ is the phason velocity, $m^\ast$ is the effective mass of electrons and $\hbar$ is the reduced Planck constant. We first consider an isolated ICDW ring with $A=0$. Assuming that $N(\varepsilon_\mathrm{F})$ is equivalent to the density of states of electrons on a one dimensional line, that is, $N(\varepsilon_\mathrm{F})=1/(\pi\hbar v_\mathrm{F})$, we have $
N_0=\hbar v_\mathrm F/(2\pi c_0^2)$. An incommensurate CDW (ICDW) can slide freely because of spatial translation symmetry, so the dynamics of an ICDW is understood by its phase $\theta$. We further assume the rigid-body model of ICDW, i.e. the ICDW ring does not deform locally and the phase $\theta(x,t)=\theta(t)$ is independent of position. Then, the Lagrangian, the canonical angular momentum and the Hamiltonian of the ICDW ring are, respectively
\begin{align}
L_0(\dot\theta)&=\int_0^{2\pi R}dx\mathscr L_0(\dot\theta)=\frac{I}{2}\dot\theta^2,
\label{Classical_Phase_Lagrangian}
\\
\pi_\theta(\dot\theta)&=\frac{\partial L_0(\dot\theta)}{\partial \dot\theta}=I\dot\theta,
\label{momentum}
\\
H_0(\pi_\theta)&=\pi_\theta\dot\theta-L_0(\dot\theta)=\frac{\pi_\theta^2}{2I}
\label{Classical_Phase_Hamiltonian}
\end{align}
where $\dot\theta=d\theta/dt$ and $I=\hbar R v_\mathrm{F}/ c_0^2$ is the moment of inertia. We note that \eqref{Classical_Phase_Lagrangian}, \eqref{momentum}, and \eqref{Classical_Phase_Hamiltonian} are time independent.
\subsection{Quantization of {an Isolated} ICDW Ring}
Next, we quantize the ICDW ring system. We show that a quantized ICDW ring possesses an inherent oscillation which originates from the uncertainty principle. Let $\hat H_0=\hat\pi_\theta^2/(2I)$ and $\hat\pi_\theta$ be the Hamiltonian operator and angular momentum operator of the ICDW ring, respectively. The macroscopic quantum state $\psi\in\mathscr H$ is defined in the Hilbert space $\mathscr H$ of positive square-integrable functions with the periodic boundary condition $\psi(\theta+2\pi)=\psi(\theta)$. The canonical commutation relation $[\hat\theta,\hat\pi_\theta]=i\hbar$ is not satisfactory because $\hat\theta$ is a multi-valued operator and is not well-defined. Ohnuki and Kitakado \cite{Ohnuki} resolved this difficulty by using the unitary operator $\hat W$ and the self-adjoint angular momentum operator $\hat \pi_\theta$ defined by
\begin{equation*}
\braket{\theta|\hat W|\psi}=e^{i\theta}\psi(\theta),\qquad
\braket{\theta|\hat \pi_\theta|\psi}=-i\hbar\frac{\partial\psi(\theta)}{\partial\theta}
\end{equation*}
which satisfy the commutation relation on $\mathscr H$
\begin{equation}
[\hat \pi_\theta,\hat W]=\hbar\hat W.\label{Algebra_Ohnuki}
\end{equation}
$\hat H_0$ is a function of $\hat\pi_\theta$ only, hence the complete orthonormal set $\{\psi_l\}_{l=-\infty}^\infty$ of momentum eigenstates spans $\mathscr H$ and satisfy $\psi_l(\theta)=e^{il\theta}/\sqrt{2\pi}$. The eigenvalues of $\hat\pi_\theta$ are quantized with $\braket{\psi_l|\hat\pi_\theta|\psi_l}=l\hbar,l\in\mathbb Z$. 
$\hat W$ and $\hat W^\dagger$ are ladder operators which satisfy $\hat W\psi_l=\psi_{l+1}$ and $\hat W\psi_l=\psi_{l-1}$. Therefore, \eqref{Algebra_Ohnuki} is the one dimensional version of the well known angular momentum algebra \cite{Sakurai}. Time evolution is introduced via the Heisenberg picture: $\hat \pi_\theta(t)=e^{i\hat H_0t/\hbar}\hat \pi_\theta e^{-i\hat H_0t/\hbar}$ and $\hat W(t)=e^{i\hat H_0t/\hbar}\hat W e^{-i\hat H_0t/\hbar}$. $\hat \pi_\theta$ commutes with $\hat H_0$, so $\hat \pi_\theta(t)=\hat \pi_\theta$. From the commutation relation \eqref{Algebra_Ohnuki} we obtain the following solutions of $\hat W(t)$:
\begin{equation}
\hat W(t)=e^{it\hat\pi_\theta/I}\hat We^{-\frac{it}{2\mu}}=\hat We^{it\hat\pi_\theta/I}e^{\frac{it}{2\mu}}\label{TimeDependentW}
\end{equation} 
where $\mu=I/\hbar$. {The two different expressions in \eqref{TimeDependentW} arise from the noncommutativity between $\hat W$ and $e^{it\hat\pi_\theta/I}$. }For a QTC we need a periodic expectation value at the ground state. So, we define the time dependent charge density operator
\begin{equation}
{\hat n(x,t)=n_0+\frac{n_1}{2}\left(e^{2ik_\mathrm F x}\hat W(t)+\mathrm{h.c.}\right)}\label{chargedensityoperator}
\end{equation}
and 
replace the classical charge density \eqref{chargedensity} by the expectation value 
\begin{equation}
n(x,t)=\braket{\hat n(x,t)}.
\end{equation}
Any states in $\mathscr H$ must be a linear superposition of $\{\psi_l\}$, that is $\psi=\sum_{l\in\mathbb Z}c_l\psi_l$ provided $\sum_{l}|c_l|^2=1$. Therefore, the expectation values $\braket{\hat W(t)}$ and $\braket{\hat n(x,t)}$ {are periodic with} period $P=4\pi\mu$ for any state $\psi$:
\begin{align}
\begin{split}
&\braket{\hat W(t)}{=\frac{1}{2}\text{tr}[\hat W(t)\hat\rho+\hat\rho\hat W(t)]}
\\
&=\frac{1}{2}\!\int_{-\pi}^\pi\! d\theta e^{i\theta}[e^{\frac{it}{2\mu}}\rho(\theta+ t/\mu,\theta)\!+\!\rho(\theta,\theta- t/\mu)e^{-\frac{it}{2\mu}}].\end{split}
\label{WExpectationValue}
\end{align}
From the two different expressions of $W(t)$ in \eqref{TimeDependentW} we can define the Weyl form of the commutation relation \cite{WeylCCR}
\begin{align*}
\hat We^{it\hat\pi_\theta/I}=e^{it\hat\pi_\theta/I}\hat We^{-\frac{it}{\mu}}\label{WeylCCR}
\end{align*}
hence the phase $\frac{t}{2\mu}$ and the periodic oscillation with period $4\pi\mu$ is a manifestation of the uncertainty principle. For an alternative explanation, let us consider an electron with effective mass $m^\ast$ confined in a finite space with volume $L\sim 2R$. From the uncertainty principle, the momentum {uncertainty} of this particle is $\Delta p\sim\hbar/(4R)$. This means that the particle's wave packet expands with velocity $v=\Delta p/m^\ast\sim\hbar/(4m^\ast R)$. Then, because of the periodic boundary condition, the physical quantity $W=e^{i x/\lambda}$ is periodic with period $P=\lambda/v\sim4\pi m^\ast R/(\hbar k_\text F)=4\pi m^\ast R/mv_\text F=4\pi Rv_\text F/c_0^2=4\pi\mu$. So, the origin of the periodicity is (i) the macroscopic wave function of the ICDW ring diffuses due to the uncertainty principle then (ii) $W=e^{i\theta}$ oscillates periodically.
However, the oscillation in \eqref{WExpectationValue} is not observable at the ground state $\hat\rho_0\equiv\ket{\psi_0}\bra{\psi_0}$ because $\braket{\theta|\hat\rho_0|\phi}=\frac{1}{2\pi}$ and the $\theta$ integral vanishes. {Therefore, the ground state of an isolated ICDW ring is not yet a time crystal because of superposition.}
\section{COUPLING TO ENVIRONMENT}
\label{With_Dissipation}
Now, suppose that the ICDW ring starts to interact with its surrounding enviromnent at $t=0$. Then, we expect decoherence of the phase $\theta$. This interaction is modeled using the Caldeira-Leggett model \cite{Caldeira} which is a model quantum Brownian motion. It describes a particle coupled to its environment. This environment is described as a set of non-interacting harmonic oscillators. {First, the classical solution $\theta(t)$ is calculated to study the dynamics of the ICDW ring. Next, this system is quantized to calculate the amplitude of the charge density expectation value $\braket{\hat n(x,t)}$}.

\subsection{Classical Theory of ICDW Ring With Environment}
Let us consider the following Lagrangian of an ICDW ring threaded by a fluctuating magnetic flux
\begin{equation}
\tilde L(\dot\theta,\mathbf q,\dot{\mathbf q})=\frac12I\dot\theta^2+A(\mathbf q)\dot\theta+\sum_{j=1}^{\mathcal N}\left(\frac12m\dot q_j^2-\frac12 m\omega_j^2q_j^2\right)
\label{ClassicalLagrangian1}
\end{equation}
where $q_j$ are the normal coordinates of the fluctuation and $\tilde\pi_\theta=\partial \tilde L/\partial\dot\theta$ and $p_j=\partial \tilde L/\partial\dot q_j$ are the canonical momenta of the ICDW ring and the environment, respectively. The fluctuating magnetic flux is given by
\begin{equation*}
A(\mathbf q)=\sum_{j=1}^\mathcal N cq_j.
\end{equation*}
Classically, this magnetic flux will randomly changes the phase and the \textit{mechanical} angular  momentum $I\dot\theta$ of the ICDW ring due to electromotive force.
An equivalent Lagrangian obtained by a Canonical transformation is
\begin{align}
\begin{split}
L(\dot\theta,\mathbf R,\dot{\mathbf R})&=\frac12I\dot\theta^2\!+\!\frac{m}{2}\sum_j^\mathcal N\!\left[\!\dot R_j^2(\theta)\!-\!\omega_j^2\left(\!\!R_j(\theta)-\frac{C_j\theta}{m\omega_j^2}\! \right)^2\right]
\label{ClassicalLagrangian2}
\end{split}
\end{align}
which is the Lagrangian of a bath of field particles $R_j$ coupled to the phase $\theta$ by springs. 
$R_j(p_j,\theta)=-\frac{p_j-c\theta}{m\omega_j}$, $P_j(q_j)=m\omega_jq_j$ and $C_j=c\omega_j$. \eqref{ClassicalLagrangian1} and \eqref{ClassicalLagrangian2} are precisely the kinds of Lagrangian considered by Caldeira and Leggett, so we can use the results in \cite{Caldeira} but with slight modifications due to the periodicity of the ring. 
\subsection{{Classical Solution}}
\begin{figure}
\includegraphics[width=0.8\linewidth]{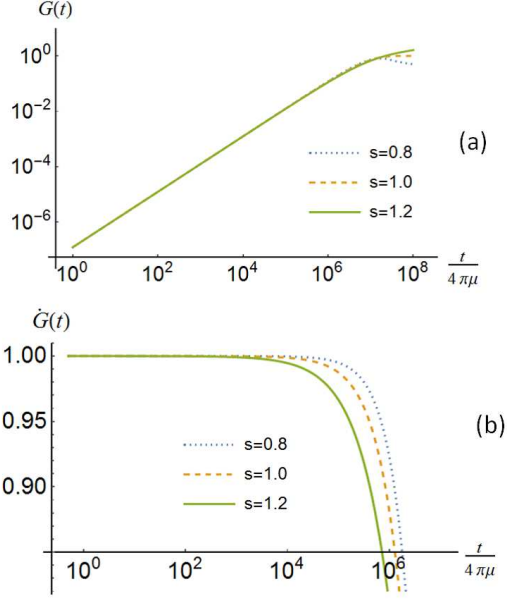}
\caption{These plots are shown with $g_s=1$ Hz$^{2-s}$ and $\mu= 10^{-8}$ sec. (a)The fundamental solution $G(t)$ of a classical ICDW ring coupled to its environment is shown for sub-ohmic ($s<1$), ohmic $(s=1)$ and super-ohmic ($s>1$) damping. Note that $G(t)\approx t$ for $t$ less than some time scale $\tau_{\text{damp},s}$. The time $t$ is normalized by $4\pi\mu$, so the horizontal axis gives the number of ``lattice points". (b)The derivative of the fundamental solution, $\dot G(t)$, heuristically describes the velocity of the ICDW ring. Note that $\dot G(t)\approx 1$ for $t\ll \tau_{\text{damp,s}}$.}
\label{ClassicalMotion}
\end{figure}

The equation of motion of $\theta$ obtained from the Lagrangian \eqref{ClassicalLagrangian2} is the generalized Langevin equation\cite{Hanggi1997}
\begin{equation}
I\ddot\theta(t)+2\int_0^td\tau\alpha_\text I(t-\tau)\theta(\tau)=\xi(t)\label{GLEModified}
\end{equation}
with the dissipation kernel $\alpha_\text I(t-\tau)$, the memory function $\gamma(t-\tau)$ and the classical fluctuating force $\xi(t)$ defined by
\begin{align*}
\alpha_{\text I}(t-\tau)&=I\gamma(0)\delta(t-\tau)+\frac{I}{2}\frac{d}{dt}\gamma(t-\tau),\label{alphaIDef}
\\
\gamma(t-\tau)&=
\sum_{j=1}^{\mathcal N}\frac{C_j^2}{Im\omega_j^2}\cos\omega_j(t-\tau),
\\
\xi(t)&=\sum_{j=1}^{\mathcal N}C_j\left[R_j(0)\cos\omega_jt+\frac{P_j(0)}{m\omega_j}\sin\omega_jt\right].
\end{align*}
The correlation function {of the classical force is} given by the noise kernel
\begin{align*}
\braket{\xi(t)\xi(\tau)}_\text{env}&=\hbar\alpha_{\text R}(t-\tau)
\\
\alpha_\text R(t-\tau)&=\sum_j^\mathcal N\frac{C_j^2}{2m\omega_j}\coth\left(\frac{\hbar\omega_j}{2k_\mathrm{B}T}\right)\cos\omega_j(t-\tau)
\end{align*}
where the average $\braket{\cdot}_\text{env}$ is taken with respect to the environment coordinate at equilibrium.
It is convenient to define the spectral density function
\begin{equation}
\mathcal J(\omega)=\frac{\pi}{2}\sum_j^\mathcal N\frac{C_j^2}{m_j\omega_j}\delta(\omega-\omega_j)
\label{J_def}
\end{equation}
and assume the power law spectrum
$\mathcal J(\omega)=I g_s\omega^s$ \cite{Grabert1988115,*Schramm1987} with a cutoff frequency $\Omega$ and $0<s<2$. Then, $\alpha_\text R(t-\tau)$ can be written
\begin{align*}
\alpha_\text R(t-\tau)&=\frac{Ig_s}{\pi}\int_0^\Omega \omega^s\coth\left(\frac{\hbar \omega}{2k_\mathrm BT}\right)\cos\omega(t-\tau)d\omega.\label{alphaRIntegral}
\end{align*}
The classical solution of $\theta$ is
\begin{equation}
\theta(t)=G(t)\dot\theta(0)+\dot G(t)\theta(0)+\frac{1}{I}\int_0^td\tau G(t-\tau)\xi(t)
\end{equation}
with the fundamental solution
\begin{equation*}
G(t)=\mathcal L^{-1}\left[\frac{1}{z^2+z\hat\gamma(z)}\right](t).
\end{equation*}
where $\mathcal L^{-1}$ is the inverse Laplace transform. The Laplace transform of the memory function $\gamma(t)$ can be written \cite{Weiss}
\begin{equation*}
\hat\gamma(z)=\omega_s^{2-s}z^{s-1},\qquad \omega_s=\left(\frac{g_s}{\sin\frac{\pi s}{2}}\right)^{1/(2-s)}
\end{equation*}
and $G(t)$ takes the form of a generalized Mittag-Leffler function $E_{\alpha,\beta}(x)=\sum_{k=0}^\infty\frac{x^k}{\Gamma(\alpha k+\beta)}$:
\begin{align*}
G(t)=tE_{2-s,2}[-(\omega_s t)^{2-s}].
\end{align*}
For ohmic damping with $s=1$ and $\hat \gamma(z)=g_1\equiv2\gamma$, we obtain
\begin{equation}
G(t)=\frac{1-e^{-2\gamma t}}{2\gamma}.\label{ClassicalMotionOhmic}
\end{equation}
{$G(t)$ and $\dot G(t)$ are shown in FIG. \ref{ClassicalMotion}. We note that $\dot G(t)\approx 1$ for $t$ less than some damping time scale $\tau_{\text{damp},s}$. In other words, the fluctuating magnetic flux does not affect the dynamics of an ICDW ring for $t<\tau_{\text{damp},s}$ and $e^{i\theta(t)}$ oscillates periodically with period $P\approx 4\pi\mu$. Next, we quantize the ICDW ring + environment system to show that this oscillation is observable for a finite time $\tau_Q$ and form an effective QTC as a metastable state.}
\subsection{Quantization {of ICDW Ring Coupled to Environment}}
{The ICDW ring $+$ environment system is} quantized using the commutation relations $[\hat{\tilde\pi}_\theta,\hat W]=\hbar\hat W$ and $[\hat q_j,\hat p_k]=i\hbar\delta_{jk}$. Define the orthonormal position state $\ket{\mathbf q}=\prod_{i=1}^\mathcal{N}\ket{q_i}$ and the orthonormal momentum state $\ket{\mathbf p}=\prod_{i=1}^\mathcal{N}\ket{p_i}$ such that
\begin{equation}
\braket{\mathbf q|\hat q_j|\psi}=q_j\braket{\mathbf q|\psi},\qquad \braket{\mathbf q|\hat p_j|\psi}=-i\hbar\frac{\partial}{\partial q_j}\braket{\mathbf q|\psi},
\end{equation}
and the inner product of $\ket{\mathbf q}$ and $\ket{\mathbf p}$ is defined as $\braket{\mathbf q|\mathbf p}=\frac{1}{\sqrt{2\pi\hbar}^\mathcal{N}}\exp\left(\frac{i}{\hbar}\mathbf q\cdot\mathbf p\right)$. The periodic boundary condition of the ring implies that $\braket{\theta+2\pi n|\tilde\pi_\theta}=\braket{\theta|\tilde\pi_\theta}$ for some integer $n$, hence the angular momentum eigenstates are quantized: $\braket{\psi_l|\hat{\tilde\pi}_\theta|\psi_l}=l\hbar, l=0,\pm,\pm2,\dots$, $\ket{\tilde\pi_\theta}=\hbar^{-1/2}\ket{\psi_l}$, $\braket{\theta|\psi_l}=\frac{1}{\sqrt{2\pi}}e^{il\theta}$. 
Moreover, one can easily show that 
\begin{equation}
\braket{\theta,\mathbf p|\tilde\pi_\theta,\mathbf q}=\braket{\theta,\mathbf R(\theta)|\pi_\theta,\mathbf P}.
\end{equation}
$\hat W$ is independent of the environmental coordinate. So, the expectation value of $\hat W$  is
\begin{align*}
\braket{\hat W(t)}
&=\frac{1}{2}\int_{-\pi}^\pi d\theta_\text f\int_{-\pi}^\pi d\phi_\text f \rho(\theta_\text f,\phi_\text f,t)\braket{\phi_\text f|\hat W|\theta_\text f}
\\
&+\frac{1}{2}\int_{-\pi}^\pi d\theta_\text f\int_{-\pi}^\pi d\phi_\text f\braket{\theta_\text f|\hat W|\phi_\text f}\rho(\phi_\text f,\theta_\text f,t).
\end{align*}
where the reduced density matrix of the ICDW ring is (see \textit{Appendix} \ref{CLS1})
\begin{align}
\begin{split}
\rho(\theta_\text f,\phi_\text f,t)&=\int_{-\pi}^{\pi}d\theta_\text i\int_{-\pi}^{\pi}d\phi_\text i\sum_{l_1,l_2\in\mathbb Z}\rho(\theta_\text i,\phi_\text i,0)
\\
&\times 
J(\theta_\text f+2\pi l_1,\phi_\text f+2\pi l_2,t;\theta_\text i,\phi_\text i,0).
\label{rho}
\end{split}
\end{align}
The exact form of $J(\theta_\text f,\phi_\text f,t;\theta_\text i,\phi_\text i,0)$ 
for ohmic dissipation $s=1$ was calculated in \cite{Caldeira}. For general damping with arbitrary $s$ the computation of the reduced density matrix is essentially equivalent to \cite{Caldeira} and we obtain
\begin{align}
J(\theta_\text f,\phi_\text f,t;\theta_\text i,\phi_\text i,0)&=F^2(t)\exp\left(\frac{i}{\hbar}S[\varphi^+_\text{cl},\varphi^-_\text{cl}]-\Gamma[\varphi^-_\text{cl}]\right).\label{JClassical}
\end{align}
$\varphi^+_\text{cl}=\frac12(\theta_\text{cl}+\phi_\text{cl})$ and $\varphi^-_\text{cl}=\phi_\text{cl}-\theta_\text{cl}$ are the classical coordinates obtained from the Euler-Lagrange equation
\begin{align}
I\ddot\varphi_\text{cl}^-(u)+2\int_u^td\tau\varphi_\text{cl}^-(\tau)\alpha_\text I(\tau-u)=0,\label{eqn1}
\\
I\ddot\varphi^+_\text{cl}(u)+2\int_0^ud\tau\varphi^+_\text{cl}(\tau)\alpha_\text I(u-\tau)=0\label{eqn2}
\end{align}
whose solution are given in terms of boundary conditions 
$\varphi^\pm_\text i=\varphi_\text{cl}^\pm(0),\varphi^\pm_\text f=\varphi_\text{cl}^\pm(t)$:
\begin{align}
\varphi^+_\text{cl}(u)&=\kappa_i(u;t)\varphi^+_\text i+\kappa_f(u;t)\varphi^+_\text f,
\label{varphisolfinal}
\\
\varphi^-_\text{cl}(u)&=\kappa_i(t-u;t)\varphi^-_\text f+\kappa_f(t-u;t)\varphi^-_\text i,
\label{varphiprimesolfinal}
\\
\kappa_i(u;t)&=\dot G(u)-\frac{\dot G(t)}{G(t)}G(u),
\qquad
\kappa_f(u;t)=\frac{G(u)}{G(t)}.
\end{align}
The classical action and the noise action are given by
\begin{align*}
S[\varphi^+_\text{cl},\varphi^-_\text{cl}]
&=S_\text{cl}(\varphi^+_\text f,\varphi^-_\text f,t;\varphi^+_\text i,\varphi^-_\text i,0)
\\
&=-I[\dot\varphi^+_\text{cl}(t)\varphi^-_\text f-\dot\varphi^+_\text{cl}(0)\varphi^-_\text i],
\\
\Gamma[\varphi_\text{cl}^-]&=\Gamma_\text{cl}(\varphi_f^-,t;\varphi_i^-,0)
\\
&=\frac{1}{2\hbar}\int_0^td\tau\int_0^td\tau'\varphi^-(\tau)\alpha_\text R(\tau-\tau')\varphi^-(s).
\end{align*}
$F^2(t)$ is a normalization function such that $\text{tr}[\rho(t)]=\braket{1}=1$. The winding numbers $l_1$ and $l_2$ can be absorbed into the $\theta_\text f$ and $\phi_\text f$ integrals, respectively, by changing the domain of $\theta_\text f$ and $\phi_\text f$ from $S^1$ to $\mathbb R^1$. Then, taking care of the non-Hermiticity of $\hat W$, we obtain
\begin{align}
\braket{\hat W(t)}&=\frac{r_1^+(t)+r_1^-(t)}{r_2^+(t)+r_2^-(t)},\label{WExpGeneral}
\\
r_1^+&=\int_{-\pi}^{\pi}d\theta_\text i\sum_{n\in\mathbb S_1(\theta,t)}\rho(\theta_\text i-f_1(t),\theta_\text i,0)\nonumber
\\
&\times e^{-in\pi-i\mu \theta \dot f_1(t)+\frac{i \mu}{2}  f_1(t) \dot f_1(t)-\Gamma_\text{cl}(2\pi n,t;f_1(t),0)},\nonumber
\\
r_1^-&=\int_{-\pi}^{\pi}d\theta_\text i\sum_{n\in\mathbb S_1(\theta,t)}\rho(\theta_\text i,\theta_\text i+f_1(t),0)\nonumber
\\
&\times e^{-in\pi-i\mu \theta \dot f_1(t)-\frac{i \mu}{2}  f_1(t) \dot f_1(t)-\Gamma_\text{cl}(2\pi n,t;f_1(t),0)},\nonumber
\\
r_2^+&=\int_{-\pi}^{\pi}d\theta_\text i\sum_{n\in\mathbb S_2(\theta,t)}\rho(\theta_\text i-f_2(t),\theta_\text i,0)\nonumber
\\
&\times e^{-i\mu \theta \dot f_2(t)+\frac{i\mu}{2}  f_2(t)\dot f_2(t)-\Gamma_\text{cl}(2\pi n,t;f_2(t),0)},\nonumber
\\
r_2^-&=\int_{-\pi}^{\pi}d\theta_\text i\sum_{n\in\mathbb S_2(\theta,t)}\rho(\theta_\text i,\theta_\text i+f_2(t),0)\nonumber
\\
&\times e^{-i\mu \theta \dot f_2(t)-\frac{i\mu}{2}  f_2(t)\dot f_2(t)-\Gamma_\text{cl}(2\pi n,t;f_2(t),0)}.\nonumber
\end{align}
with $f_1(t)=2\pi n \dot G(t)-G(t)/\mu$, $f_2(t)=2\pi n \dot G(t)$, $\mathbb S_1(\theta,t)=\{n\in\mathbb Z|-\pi<\theta+f_1(t)<\pi\}$, and $\mathbb S_2(\theta,t)=\{n\in\mathbb Z|-\pi<\theta+f_2(t)<\pi\}$.
This is the most general form of the expectation value of $\hat W$ for an ICDW ring coupled to its environment.
Although the derivation of \eqref{WExpGeneral} is exact, it is not very insightful, so we make some approximations. 
\subsection{Early Time {Approximation}}
Let us consider the classical solution (22) with $\varphi^+_\text{cl}=\theta_\text f$ and ohmic damping \eqref{ClassicalMotionOhmic}, then we see immediately that $\dot\theta(t)\sim \dot\theta(0)e^{-2\gamma t}$. Therefore, we are interested in the range $t\ll1/(2\gamma)\equiv\tau_{\text{damp},1}$. For general dissipation, we can see from FIG. \ref{ClassicalMotion} that there exist a time scale $\tau_{\text{damp},s}$ such that $G(t)\approx t$, $\dot G(t)\approx 1$ for $t\ll\tau_{\text{damp},s}$. Then, writing $t=2\pi I(m+a)/\hbar$ for an integer $m$ and $0<a<1$, we have $\mathbb S_1(\theta<-2a\pi,t)=\{m+1\}$, $\mathbb S_1(\theta>-2a\pi,t)=\{m\}$, and $\mathbb S_2(\theta,t)=\{0\}$. Therefore, we conclude that $\mathbb S_1(\theta,t)$ is approximately the $(m/2)^{\text{th}}$ lattice point.  Then, using $\dot f_1(t)\approx -\dot G(t)\hbar/I$ we obtain the approximate form 
\begin{align*}
\braket{\hat W(t)}&\approx\int_{-\pi}^{\pi}d\theta_\text i\sum_{n\in\mathbb S_1(\theta,t)}\rho(\theta_\text i,\theta_\text i-G(t)/\mu,0)
\\
&\times\exp\left\{i\theta_\text i\dot G(t)-i\frac{G(t)\dot G(t)}{2\mu}-\Gamma_\text{cl}(2\pi n,t;f_1(t),0)\right\}
\\
&+\int_{-\pi}^{\pi}d\theta_\text i\sum_{n\in\mathbb S_1(\theta,t)}\rho(\theta_\text i+G(t)/\mu,\theta_\text i,0)
\\
&\times\exp\left\{i\theta_\text i\dot G(t)+i\frac{G(t)\dot G(t)}{2\mu}-\Gamma_\text{cl}(2\pi n,t;f_1(t),0)\right\}.
\end{align*}
For $t\ll\tau_{\text{damp},s}$ we have $\varphi^-_\text{cl}(\tau)\approx \tau/\mu$ and the noise action can be written
\begin{align*}
&\Gamma_\text{cl}(2\pi n,t;f_1(t),0)\approx \Gamma_{T,s}(t)
\\
&\qquad=\frac{g_s}{2\pi \mu}\int_0^\Omega d\omega\coth\left(\frac{\hbar \omega}{2k_\text BT}\right)\Upsilon(\omega),
\\
&\Upsilon(\omega)=\omega^{s-4}(2+\omega^2t^2-2\cos\omega t-2\omega t\sin\omega t)
\end{align*}
\begin{figure}
\includegraphics[width=0.8\linewidth]{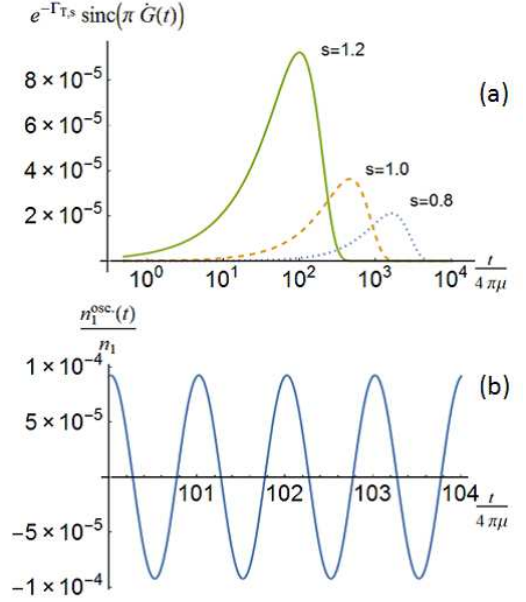}
\caption{(a) The amplitude of the charge density oscillation is shown for $g_s=1$ Hz$^{2-s}$, $\mu=10^{-8}$sec. and $\Omega=1/\mu$. This oscillation is an effective QTC. In general, the oscillation amplitude is larger for super-ohmic($s>1$) damping but has a small life time $\tau_{Q}$. On the other hand, the oscillation amplitude is small for sub-ohmic ($s<1$) damping but has a long $\tau_Q$. (b) The charge density oscillation is shown for super-ohmic damping with $s=1.2$, $g_{1.2}=1$ Hz$^{0.8}$, $\mu=10^{-8}$sec. and $\Omega=1/\mu$. The time $t$ is normalized such that the horizontal axis gives the number of ``lattice points".}
\label{GSOsc}
\end{figure}
Taking the low temperature limit $\frac{\hbar \Omega}{2k_\text BT}\to\infty$ such that $\coth\frac{\hbar \Omega}{2k_\text BT}\to1$ we obtain 
\begin{align*}
\Gamma_{T,s}(t)
=&-\frac{g_s}{\pi \mu}\frac{\Omega ^{s-3} \left(\,
   _1F_2\left(\frac{s-3}{2};\frac{1}{2},\frac{s-1}{2};-\frac{1}
   {4} t^2 \Omega ^2\right)-1\right)}{s-3}
\\
&-\frac{g_st^2}{2\pi \mu}\frac{\Omega ^{s-1} \left(\,
   _1F_2\left(\frac{s-1}{2};\frac{1}{2},\frac{s+1}{2};-\frac{1}
   {4} t^2 \Omega ^2\right)-1\right)}{s-1}
\end{align*}
where $\,_1F_2\left(a_1;b_1,b_2;z\right)$ is the generalized hypergeometric function. Define the decoherence time $\tau_{\text{decoh},s}$ such that $\Gamma_{T,s}(\tau_{\text{decoh},s})=1$. 
Numerical analysis shows that the order of $\tau_{\text{decoh},s}$ does not change with $\Omega<1/(2\mu)$ and decreases rapidly for $\Omega>1/(2\mu)$. 
If we set $\Omega\sim1/\mu$ \st{we obtain $\tau_{\text{decoh},s}\sim g_s^{1/4}\mu ^{(s+2)/4}$ and the number of observable ``lattice points" is $N\sim g_s^{1/4}\mu ^{(s-2)/4}$. If we}\textcolor{red}{and} use the ground state $\rho_0$, then for $t\ll\tau_Q\equiv\min\{\tau_{\text{damp,s}},\tau_{\text{decoh},s}\}$ we have
\begin{align}
n(x,t)&\approx n_0+n^\text{osc.}_1(t)\cos(2k_\text F x),\label{CDWResult}\\
n^\text{osc.}_1(t)&=n_1\text{sinc}[\pi\dot G(t)]\cos\left[\frac{\dot G(t)G(t)}{2\mu}\right]e^{-\Gamma_{T,s}(t)}.
\end{align}
Equation \eqref{CDWResult} is the main result of this paper. It shows that the amplitude of an ICDW ring threaded by a (time independent) fluctuating magnetic flux oscillates for a finite time $\tau_Q$ and form an effective QTC. 
In the no-damping limit $\hat\gamma(z)\to0$ we have $G(t)\to t,G(t)\to 1$ and recover $\braket{\hat W(t)}=0$. This charge density oscillation is shown in FIG. \ref{GSOsc}.
\section{DISCUSSION}
\label{Conclusion}
\textcolor{red}{First, we elaborate our assumtion of ICDW ring. A mathematical definition of ICDW is that $\lambda/a$ is an irrational number. We note that a CDW formed on a macroscopic crystal is basically incommensurate because the wavelength of a CDW is given by $\lambda=\pi/k_\text{F}$, where the Fermi wave number $k_\text{F}$ is usually an irrational number for an arbitrary band filling. However, strictly speaking, $\lambda/a$ of a finite-size system can never be irrational. A physical condition is that the commensurability pinning energy is negligible, which is possible if $\lambda/a$ cannot be expressed as a simple fraction like 2, 5/2, etc.  In order to explain this, suppose that for some integer  $M\geq2$, $Ma/\lambda$ is an integer. In other words, the same atom-electron configuration is obtained if we move the CDW by $M$ wavelengths and $\epsilon_{k+2Mk_\text{F} }=\epsilon_k$ ($\epsilon_k$ is the energy of an electron with momentum $\hbar k$). Then, the energy required to move a CDW by a small phase $\phi$ from its equilibrium is \cite{LRA} 
\begin{equation*}
\epsilon(\phi;M)=\frac{|\Delta|^2}{\epsilon_\text{F}}  \left(\frac{e|\Delta|}{W}\right)^{M-2}\frac{M\phi^2}{2}
\end{equation*}
where $\epsilon_\text{F}$ is the Fermi energy, $W$ is the band width, $|\Delta|$ is the CDW gap width and $e$ is the elementary charge in natural units. We see that $\epsilon(\phi;M)$ approaches zero rapidly for large $M$ as the distinction between rational and irrational numbers becomes academic. For example, for a ring crystal with $N_a$ atoms and $N_\lambda$ CDW wavelengths, we obtain $\lambda/a=N_a/N_\lambda$.  Therefore, for a large $N_a$ and a large $N_\lambda$, $M$ can always be arbitrary large (order of $N_a$), hence the commensurability energy is completely negligible. In fact, we can experimentally make sub-micrometer scale ICDW rings such that, we expect, $N_a$ and $N_\lambda$ are not so large but $M$ is very large such that commensurability energy is negligible. }

\textcolor{red}{
Usually, superposition of ICDWs with different phases is not observed because of impurity pinning. However, the probability of impurity decreases with decreasing radius. Commensurability effect may become significant for small radius (more precisely, for some small $M$). But, the origin of the “time crystal periodicity” in our model is the uncertainty principle, which appears as a collective fluctuation of the ICDW phase. If commensurability effect becomes significant, then the ICDW phase is expected to fluctuate periodically around some phase $\theta=\theta_0$ determined by the ions’ position. This fluctuation is expected to become apparent and oscillate periodically by coupling to environment (fluctuating magnetic flux).}

\textcolor{red}{Next, we discuss the presence of an upper bound and a lower bound for the radius of the ICDW ring in our model. Decoherence induced breaking of time translation symmetry occurs, in principle, only in mesoscopic systems: There is an upper bound for the CDW radius determined by the coupling strength  $\gamma=\omega_s/2$ and a lower bound given by the CDW wavelength $\lambda$. We will first calculate the upper bound by replacing the environment with an equivalent LC circuit. Here, we focus on Ohmic damping because an approximate form of $\gamma$ can be calculated explicitly. In general, the coupling strength $\gamma$ depends on the CDW radius $R$.  In order to explicitly see the $R$ dependence, suppose that the fluctuating magnetic flux in our model comes from an external coil connected to a series of parallel capacitors. Then, the Lagrangian of the CDW + environment system is
\begin{equation*}
L_\text{circuit}=L_0+MI_{CDW} \sum_{j=1}^\mathcal{N}I_j +\sum_{j=1}^\mathcal{N}\frac m2(I_j^2-\omega_j^2 Q_j^2 ).
\end{equation*}
$L_0$ is the Lagrangian of an isolated ICDW ring, $I_{CDW}=e\dot\theta/\pi$ is the CDW current induced by the fluctuating flux, $Q_j$ is the net charge on the capacitor $j$ with capacitance $\mathscr C_j$, $m$ is the inductance of the coil, $\omega_j=1/\sqrt{m\mathscr C_j}$, and $M$ is the mutual inductance between the coil and the CDW. We immediately see that $L_\text{circuit}$ is equivalent to the Lagrangian in equation \eqref{ClassicalLagrangian1} but with the interaction Lagrangian replaced by $\dot \theta\sum_{j=1}^\mathcal N \frac{Me}{\pi} \dot Q$.  Therefore, define $p_j=mI_j$,  $C_j=Me\omega_j^2/\pi$ and we obtain the Lagrangian in equation \eqref{ClassicalLagrangian2} after the canonical transformation
\begin{equation*}
L=L_\text{circuit}-\frac{d}{dt}\sum_j^\mathcal{N}(MI_{CDW}+p_j) Q_j
\end{equation*}
with $P_j=m\dot R_j=m\omega_j Q_j$ and $R_j=-(p_j-MI_{CDW})/(m\omega_j)$. Assume that the radius of the coil $r_\text{coil}$ is much larger than the CDW radius $R$ and that the coil and the CDW are concentric. Then, we obtain  $M=(\mu_0 \pi R^2)/(2r_\text{coil} )$.  Next, integrate \eqref{J_def} with respect to $\omega$ from $\omega=0$ to $\omega=\Omega=1/\mu$ and obtain $\gamma=\beta R$, $\beta=(\pi \mu_0^2 e^2 \rho c_0^6  )/(32\hbar r_\text{coil}^2 m v_F^3 )$.
Here, $\mu_0$ is the permeability of free space and $\rho$ is the density of states defined by $\sum_j\to\int d\omega\rho$. If we assume that $\rho\sim\Omega$ like in the Caldeira-Leggett model, then the order of $\beta$ may change depending on the parameters of the CDW and the parameters of the coil, but $\gamma=\beta R$ is usually smaller than 1Hz.
Now, if the radius of the CDW ring is too large, then periodicity does not appear because the oscillation period exceeds the lifetime $\tau_Q$ of the time crystal. The upper bound of the CDW radius R is given by the condition $N>1$, where $N=\min\{\tau_{\text{damp},s},\tau_{\text{deph},s}\}/(4\pi\mu)$ is the number of oscillations. For Ohmic damping ($s=1$) we have the approximate form $\tau_{\text{damp},1}=\gamma^{-1}$ and $\tau_{\text{deph},1}=\sqrt{\mu\gamma^{-1}}=\sqrt{\mu\gamma}\tau_{\text{damp},1}$. 
Let $\mu\gamma<1$, i.e. $\tau_{\text{damp},1}>\tau_{\text{deph},1}$ (which is a valid assumption because a typical value for $\mu$ with radius $10^{-6}$m is $10^{-6}$s and $\gamma$ is usually smaller than 1Hz), then, the upper bound to observe more than one oscillation is $R<\frac{c_0}{4\pi\sqrt{v_F\beta}}$ which is typically 1mm. 
There is also a lower bound determined by the CDW wavelength $\lambda$: The radius should be large enough to define a Fermi surface. This condition is given by $p_\text F=\hbar\pi/\lambda\gg\hbar/R$. In other words, $R$ should be much larger than $\lambda$.
}

Finally, we discuss how our model can be tested experimentally and discuss how our results may be applicable to other physical systems. Ring-shaped crystals and ICDW ring crystals (such as monoclinic TaS$_3$ ring crystals and NbSe$_3$ ring crystals) have been produced and studied by the Hokkaido group \cite{ring,Matsuura,ABCDW}. Therefore, our model can be tested provided ring crystals with almost no defects and impurities can be produced. The oscillation in \eqref{CDWResult} implies that the local charge density of the ICDW ring oscillates with frequency $\omega=\hbar/2I$. For a ring with diameter $2R=1\mu$m, $v_\text F/c_0= 10^3$ and $v_\mathrm{F}=10^5$m/s, we have $\omega=10^8Hz.$ The time dependence of the charge density modulation can be measured using scanning tunneling microscopy (STM) \cite{Ichimura} and/or using narrow-band noise with vanishing threshhold voltage \cite{IDO}. 

We recall that the origin of the quantum oscillation in our model is the uncertainty principle. If we were to consider a single particle with mass $m^\ast$ confined on a ring with radius $R$, then the particle's wave packet expands with a velocity $v\sim \hbar/(4m^\ast R)$ and $e^{i\theta}$ oscillates with period $P=2\pi R/v$. However, a charge density wave has an internal periodicity given by the wavelength $\lambda$. Then the period of oscillation is $P\sim\lambda/v$. Therefore, because an ICDW ring is described by a macroscopic wave function with internal periodicity, the number of lattice points $N$ in our model is numerous. And yet, the periodicity of $\hat W(t)$ seems to be universal for any wave functions on $S^1$ (ring system). Therefore, our results may be applicable to earlier models of QTC such as \cite{QTC,Wigner,Josephson} and annular Josephson junctions \cite{AnnularJJ}. Moreover, it was shown vely recently in \cite{Haffner} that the ground state of a $^{40}$Ca$^+$ ring trap possesses rotational symmetry as the number of ions is decreased. Our results predict that quantum oscillations may appear in such ring traps with the appropriate set up.
We also recall that Volovik's proposal of metastable effective QTC \cite{Volovic} is not restricted to ring systems. Therefore, time translation symmetry breaking by \textit{decoherence} may occur in other systems coupled to time-independent environment, and without a periodic driving field. 
We also expect that many other incommensurate systems such as incommensurate spin density waves \cite{Gruner2}, incommensurate mass density waves \cite{PhysRevLett.40.1507,PhysRevB.20.751,PhysRevB.71.104508}, or possibly some dielectrics that exhibit incommensurate phases \cite{blinc1986incommensurate}, may be used to test our results and to model  QTC without spontaneous symmetry breaking.
\acknowledgments
We thank Kohichi Ichimura, Toru Matsuura, Noriyuki Hatakenaka, Avadh Saxena, Yuji Hasegawa, Kousuke Yakubo and Tatsuya Honma for stimulating and valuable discussions.
\bibliographystyle{apsrev4-1}
\onecolumngrid
\bibliography{20170829APS_With_Referee_Reply}
\appendix
\section{QUANTUM BROWNIAN MOTION ON $S^1$}
\label{CLS1}
\subsection{Reduced Density Matrix On $S^1$}
Consider an ICDW ring $A$ coupled to its environment $B$. 
Let $\rho_A$ and $\rho_B$ denote the density operators of $A$ and $B$, respectively. Let $\theta$ and $\phi$ be angular coordinates on the ring system $A$ and let $\mathbf p=\{p_k:k=1,\mathcal N\}$ and $\mathbf s=\{s_k:k=1,\mathcal N\}$ be momentum coordinates of the bath $B$. Then, the density matrix of the coupled system can be written
\begin{align*}
\rho(\theta,\mathbf p,\phi,\mathbf s)&=
\braket{\theta , \mathbf p|\rho_{AB}(t)|\phi ,\mathbf  s}
\\
&=\int_0^{2\pi}d\theta'd\phi'\int_{-\infty}^{\infty}d\mathbf p'd\mathbf s'K(\theta ,\mathbf p,t;\theta',\mathbf p',0)K^\ast(\phi,\mathbf s,t;\phi',\mathbf s',0)\\
&\quad\times\braket{\theta',\mathbf p'|\rho_{AB}(0)|\phi',\mathbf s'}
\end{align*}
where
\begin{equation}
K(\theta,\mathbf p,t;\theta',\mathbf p',0)=\braket{\theta,\mathbf p|e^{-\frac{i}{\hbar}\hat H t}|\theta',\mathbf p'}
\label{PropagatorTheta}
\end{equation}
and
\begin{equation}
K^\ast(\phi,\mathbf s,t;\phi',\mathbf s',0)=\braket{\phi',\mathbf s'|e^{\frac{i}{\hbar}\hat H t}|\phi,\mathbf s}
\label{PropagatorPhi}
\end{equation}
are recognized as Feynman propagators if we notice that $\ket{\mathbf p}$ and $\ket{\mathbf s}$ are actually position states after canonical transformation. The propagators \eqref{PropagatorTheta} and \eqref{PropagatorPhi} can be written using path integrals by dividing the time $t$ into $N$ time steps of length $\epsilon=t/(N+1)$, $\mathbf p=\mathbf p_N$, $\mathbf p'=\mathbf p_0$, $\theta=\theta_N$ and $\theta'=\theta_0$. For $N\to\infty$ the propagator becomes
\begin{align*}
K(\theta,\mathbf p,t;\theta',\mathbf p',0)&=\Braket{\theta,\mathbf p|\lim_{\epsilon\to 0}\left(\exp-\frac{i\epsilon}{\hbar}\hat H\right)^N|\theta',\mathbf p'}\\
&=\lim_{\epsilon\to 0}\int_0^{2\pi}\left(\prod_{n=1}^{N-1}d\theta_n\right)\int_{-\infty}^\infty\left(\prod_{n=1}^{N-1}d\mathbf p_n\right)\prod_{n=1}^NK_n(\theta_n,\mathbf p_n,\epsilon;\theta_{n-1},\mathbf p_{n-1},0).
\end{align*}
The Hamiltonian operator $\hat H$ can be decomposed into the kinetic part $\hat{\mathcal K}$ and the potential part $\hat{\mathcal V}$, i.e. $\hat H=\hat{\mathcal K}+\hat{\mathcal V}$, which satisfy the eigenvalue equations
\begin{equation}
\hat{\mathcal K}\ket{\psi_l,\mathbf q}=\mathcal K(l,\mathbf P)\ket{\psi_l,\mathbf q},\quad \hat{\mathcal V}\ket{\theta,\mathbf p}=\mathcal V(\theta,\mathbf R(\theta))\ket{\theta,\mathbf p}.
\end{equation}
Then we obtain
\begin{align*}
K&(\theta_n,\mathbf p_n,\epsilon;\theta_{n-1},\mathbf p_{n-1},0)\\
&=\sum_{l_n}\int_{-\infty}^\infty d\mathbf q_n\Braket{\theta_n,\mathbf p_n|\exp\left(-\frac{i\epsilon}{\hbar}\hat{\mathcal K}\right)|\psi_{l_n},\mathbf q_n}\Braket{\psi_{l_n},\mathbf q_n|\exp\left(-\frac{i\epsilon}{\hbar}\hat{\mathcal V}\right)|\theta_{n-1},\mathbf p_{n-1}}\\
&=\sum_{l_n=-\infty}^\infty\frac{1}{2\pi}\int_{-\infty}^\infty \frac{d\mathbf q_n}{(2\pi \hbar)^{\mathcal N}}\exp\left\{-\frac{i\epsilon}{\hbar}\mathcal K(l_n,\mathbf P_n)-\frac{i\epsilon}{\hbar}\mathcal V(\theta_{n-1},\mathbf R_{n-1}(\theta_{n-1}))\right\}\\
&\quad\times\exp\left\{il_n(\theta_n-\theta_{n-1})-\frac{i}{\hbar}\mathbf q_n\cdot(\mathbf p_n-\mathbf p_{n-1}))\right\}.
\end{align*}
Next, define $A_n=\sum_jcq_{j,n}=\sum_j\frac{C_j}{m\omega_j^2}P_{j,n}$ and obtain
\begin{align*}
K&(\theta_n,\mathbf p_n,\epsilon;\theta_{n-1},\mathbf p_{n-1},0)\\
&=\sum_{l_n=-\infty}^\infty\frac{1}{2\pi}\int_{-\infty}^\infty \frac{d\mathbf q_n}{(2\pi \hbar)^{\mathcal N}}\exp\left\{-\frac{i\epsilon}{\hbar}\mathcal K(l_n,\mathbf P_n)-\frac{i\epsilon}{\hbar}\mathcal V(\theta_{n-1},\mathbf R_{n-1}(\theta_{n-1}))\right\}\\
&\quad\times\exp\left\{i(l_n-A_n/\hbar)(\theta_n-\theta_{n-1})+\frac{i}{\hbar}\mathbf P_n\cdot(\mathbf R_n(\theta_n)-\mathbf R_{n-1}(\theta_{n-1}))\right\}.
\end{align*}
The sum over $l_n$ can be replaced by a sum of integrals using the Poisson resummation formula
\begin{equation}
\sum_{l\in\mathbb Z}f(l)=\sum_{l\in\mathbb Z}\int_{-\infty}^\infty f(\zeta)\exp(2\pi l \zeta)d\zeta.
\end{equation}
Then,
\begin{align*}
K&(\theta_n,\mathbf p_n,\epsilon;\theta_{n-1},\mathbf p_{n-1},0)\\
&=\sum_{l_n=-\infty}^\infty\int_{-\infty}^\infty\frac{d\zeta_n}{2\pi}\int_{-\infty}^\infty \frac{d\mathbf q_n}{(2\pi \hbar)^{\mathcal N}}\exp\left\{-\frac{i\epsilon}{\hbar}\mathcal K(\zeta_n,\mathbf P_n)-\frac{i\epsilon}{\hbar}\mathcal V(\theta_{n-1},\mathbf R_{n-1}(\theta_{n-1}))\right\}\\
&\quad\times\exp\left\{i(\zeta_n-A_n/\hbar)(\theta_n-\theta_{n-1}+2\pi l_n)+\frac{i}{\hbar}\mathbf P_n\cdot(\mathbf R_n(\theta_n+2\pi l_n)-\mathbf R_{n-1}(\theta_{n-1}))\right\}.
\end{align*}
For our Hamiltonian of an ICDW ring threaded by a fluctuating magnetic flux we have
\begin{align*}
\mathcal K(\zeta_n,\mathbf P_n)&=\frac{\left(\zeta_n\hbar-A_n\right)^2}{2I}+\sum_j\frac{1}{2m}P_{j,n}^2\\
\mathcal V(\theta_{n-1},\mathbf R_{n-1}(\theta_{n-1}))&=\sum_j\frac{1}{2}m\omega_j^2 \left(R_{j,n-1}(\theta_{n-1})-\frac{C_j\theta_{n-1}}{m\omega_j^2}\right)^2.
\end{align*}
We note that that the potential $\mathcal V(\theta,\mathbf R(\theta))$ is rotationally invariant. That is, for an arbitrary rotation $\theta\to\theta+\delta$, the potential is $\mathcal V(\theta+\delta,\mathbf R(\theta+\delta))=\mathcal V(\theta,\mathbf R(\theta))$. 
Let $\tilde\zeta_n=\zeta_n-A_n/\hbar$. We use the fact that the phase space volume is conserved under canonical transformation.
Then,
\begin{align*}
K&(\theta,\mathbf p,t;\theta',\mathbf p',0)
\\
&=\left(\prod_{j=1}^\mathcal{N}\frac{1}{m\omega_j}\right)\lim_{\epsilon\to 0}\left(\prod_{n=1}^{N-1}\int_0^{2\pi}d\theta_n\int_{-\infty}^\infty d\mathbf R_{n}(\theta_n)\right)\left(\prod_{n=1}^N\sum_{l_n=-\infty}^\infty\int_{-\infty}^\infty\frac{d\mathbf P_n}{(2\pi \hbar)^{\mathcal N}}\frac{d\tilde \zeta_n}{2\pi}\right)
\\
&\times\exp\sum_{n=1}^N\left\{-\frac{i\epsilon}{\hbar}\left(\frac{\hbar^2}{2I}\tilde \zeta_n^2+\sum_j\frac{1}{2m}P_{j,n}^2\right)-\frac{i\epsilon}{\hbar}\mathcal V(\theta_{n-1},\mathbf R_{n-1}(\theta_{n-1}))\right\}
\\
&\times\exp\sum_{n=1}^N\left\{i\tilde\zeta_n(\theta_n-\theta_{n-1}+2\pi l_n)+\frac{i}{\hbar}\mathbf P_n\cdot(\mathbf R_n(\theta_n+2\pi l_n)-\mathbf R_{n-1}(\theta_{n-1}))\right\}\\
\end{align*}
Solve the $\tilde \zeta_n$ and $P_{j,n}$ integrals to obtain
\begin{align*}
K&(\theta,\mathbf p,t;\theta',\mathbf p',0)\\
&=\left(\prod_{j=1}^\mathcal{N}\frac{1}{m\omega_j}\right)\lim_{\epsilon\to 0}\left(\prod_{n=1}^{N-1}\int_0^{2\pi}d\theta_n\int_{-\infty}^\infty d\mathbf R_{n}(\theta_n)\right)\left(\prod_{n=1}^N\sum_{l_n=-\infty}^\infty\sqrt{\frac{I}{2\pi i\epsilon\hbar}}\sqrt{\frac{m}{2\pi i\epsilon \hbar}}^{\mathcal N}\right)\\
&\times\exp\sum_{n=1}^N\left\{\frac{i}{\hbar}\frac{I}{2\epsilon}(\theta_n-\theta_{n-1}+2\pi l_n)^2+\frac{i}{\hbar}\frac{m}{2\epsilon}(\mathbf R_n(\theta_n+2\pi l_n)-\mathbf R_{n-1}(\theta_{n-1}))^2\right\}\\
&\times\exp\sum_{n=1}^N\left\{-\frac{i\epsilon}{\hbar}\mathcal V(\theta_{n-1},\mathbf R_{n-1}(\theta_{n-1}))\right\}.
\end{align*}
The integral including $\theta_{N-1}$ is
\begin{align*}
I_{N-1}&=\sum_{l_N=-\infty}^\infty\sum_{l_{N-1}=-\infty}^\infty\sqrt{\frac{I}{2\pi i\epsilon\hbar}}\sqrt{\frac{m}{2\pi i\epsilon \hbar}}^{\mathcal N}\int_0^{2\pi}d\theta_{N-1}\int_{-\infty}^\infty d\mathbf R_{N-1}(\theta_{N-1})\\
&\times\exp\left\{\frac{i}{\hbar}\frac{I}{2\epsilon}(\theta_N-\theta_{N-1}+2\pi l_N)^2+\frac{i}{\hbar}\frac{I}{2\epsilon}(\theta_{N-1}-\theta_{N-2}+2\pi l_{N-1})^2\right\}\\
&\times\exp\left\{\frac{i}{\hbar}\frac{m}{2\epsilon}[\mathbf R_N(\theta_N+2\pi l_N)-\mathbf R_{N-1}(\theta_{N-1})]^2\right\}\\
&\times\exp\left\{\frac{i}{\hbar}\frac{m}{2\epsilon}[\mathbf R_{N-1}(\theta_{N-1}+2\pi l_{N-1})-\mathbf R_{N-2}(\theta_{N-2})]^2\right\}\\
&\times\exp\left\{-\frac{i\epsilon}{\hbar}\mathcal V(\theta_{N-1},\mathbf R_{N-1}(\theta_{N-1}))\right\}.
\end{align*}
The sum over $l_{N-1}$ can be absorbed into the $\theta_{N-1}$ integral by changing the domain of $\theta_{N-1}$ integration from $[0,2\pi)$ to $(-\infty,\infty)$. This is done by the following procedure:
\begin{enumerate}
\item transform $l_{N}$ to $\tilde l_N=l_N+l_{N-1}$ and write $\tilde \theta_{N-1}=\theta_{N-1}+2\pi l_{N-1}$ to obtain
\begin{align*}
I_{N-1}&=\sum_{l_N=-\infty}^\infty\sum_{l_{N-1}=-\infty}^\infty\sqrt{\frac{I}{2\pi i\epsilon\hbar}}\sqrt{\frac{m}{2\pi i\epsilon \hbar}}^{\mathcal N}\int_0^{2\pi}d\theta_{N-1}\int_{-\infty}^\infty d\mathbf R_{N-1}(\theta_{N-1})
\\
&\times\exp\left\{\frac{i}{\hbar}\frac{I}{2\epsilon}(\tilde\theta_N-\theta_{N-1}+2\pi \tilde l_N)^2+\frac{i}{\hbar}\frac{I}{2\epsilon}(\tilde \theta_{N-1}-\theta_{N-2})^2\right\}
\\
&\times\exp\left\{\frac{i}{\hbar}\frac{m}{2\epsilon}[\mathbf R_N(\theta_N+2\pi \tilde l_N)-\mathbf R_{N-1}(\tilde\theta_{N-1})]^2\right\}
\\
&\times\exp\left\{\frac{i}{\hbar}\frac{m}{2\epsilon}[\mathbf R_{N-1}(\tilde\theta_{N-1})-\mathbf R_{N-2}(\theta_{N-2})]^2\right\}
\\
&\times\exp\left\{-\frac{i\epsilon}{\hbar}\mathcal V(\tilde\theta_{N-1},\mathbf R_{N-1}(\tilde\theta_{N-1}))\right\}.
\end{align*}
\item Change the variable of integration $\int_0^{2\pi}d\theta_n\to\int_{2\pi n}^{2\pi(n+1)}d\tilde \theta_n$ and change the domain of $\theta_{N-1}$ integration from $[0,2\pi)$ to $(-\infty,\infty)$ and use the periodicity of $\mathcal V(\theta_{N-1},\mathbf R_{N-1}(\theta_{N-1}))$ to obtain
\begin{align*}
I_{N-1}&=\sum_{l_N=-\infty}^\infty\sqrt{\frac{I}{2\pi i\epsilon\hbar}}\sqrt{\frac{m}{2\pi i\epsilon \hbar}}^{\mathcal N}\int_{-\infty}^\infty d\theta_{N-1}\int_{-\infty}^\infty d\mathbf R_{N-1}(\theta_{N-1})\\
&\times\exp\left\{\frac{i}{\hbar}\frac{I}{2\epsilon}(\theta_N-\theta_{N-1}+2\pi l_N)^2+\frac{i}{\hbar}\frac{I}{2\epsilon}(\theta_{N-1}-\theta_{N-2})^2\right\}\\
&\times\exp\left\{\frac{i}{\hbar}\frac{m}{2\epsilon}[\mathbf R_N(\theta_N+2\pi l_N)-\mathbf R_{N-1}(\theta_{N-1})]^2\right\}\\
&\times\exp\left\{\frac{i}{\hbar}\frac{m}{2\epsilon}[\mathbf R_{N-1}(\theta_{N-1})-\mathbf R_{N-2}(\theta_{N-2})]^2\right\}\\
&\times\exp\left\{-\frac{i\epsilon}{\hbar}\mathcal V(\theta_{N-1},\mathbf R_{N-1}(\theta_{N-1}))\right\}.
\end{align*}
\end{enumerate}

Repeat this procedures for the integrals involving $\theta_{n}$ from $n=N-2$ to $n=1$, and obtain
\begin{align*}
K&(\theta,\mathbf q,t;\theta',\mathbf q',0)\\
&=\sum_{l=-\infty}^\infty\lim_{\epsilon\to 0}\int_{-\infty}^{\infty}\left(\prod_{n=1}^{N-1}d\theta_n\right)\int_{-\infty}^\infty\left(\prod_{n=1}^{N-1}d\mathbf R_{n-1}(\theta_{n-1})\right)\prod_{n=1}^N\sqrt{\frac{I}{2\pi i\epsilon\hbar}}\sqrt{\frac{m}{2\pi i\epsilon \hbar}}^{\mathcal N}\\
&\times\exp\left\{\frac{i}{\hbar}\frac{I}{2\epsilon}(\theta_n-\theta_{n-1}+2\pi l\delta_{n,N})^2\right\}\\
&\times\exp\left\{\frac{i}{\hbar}\frac{m}{2\epsilon}[\mathbf R_n(\theta_n+2\pi l\delta_{n,N})-\mathbf R_{n-1}(\theta_{n-1})]^2-\frac{i\epsilon}{\hbar}\mathcal V(\theta_{n-1},\mathbf R_{n-1}(\theta_{n-1}))\right\}.
\end{align*}
Define
\begin{align}
\mathcal A_A&=\sqrt{\frac{2\pi i\epsilon\hbar}{I}},\qquad \mathcal A_B=\sqrt{\frac{2\pi i\epsilon\hbar}{m}}^\mathcal{N},
\\
\int \mathcal{D}\theta&=\frac{1}{\mathcal A_A}\int_{-\infty}^\infty\left(\prod_{n=1}^{N-1}\frac{d\theta_n}{\mathcal A_A}\right),
\qquad
\int \mathcal{D}\mathbf R(\theta)=\frac{1}{\mathcal A_B}\int_{-\infty}^\infty\left(\prod_{n=1}^{N-1}\frac{d\mathbf R_n(\theta_n)}{\mathcal A_B}\right),
\end{align}
\begin{align}
L_n(\theta_n,\mathbf R_n(\theta_n))&=\frac{I}{2}\left(\frac{\theta_n-\theta_{n-1}+2\pi l\delta_{n,N}}{\epsilon}\right)^2\nonumber
\\
&+\frac{m}{2}\left(\frac{\mathbf R_n(\theta_n+2\pi l\delta_{n,N})-\mathbf R_{n-1}(\theta_{n-1})}{\epsilon}\right)^2-\mathcal V(\theta_{n-1},\mathbf R_{n-1}(\theta_{n-1}))
\end{align}
where
\begin{align}
L[\theta,\mathbf R(\theta)]&=\lim_{\epsilon\to 0}L_n[\theta_n,\mathbf R_n(\theta_n)]=L_0[\theta]+L_B[\theta,\mathbf R(\theta)]
\\
L_0[\theta]&=\frac{I}{2}\dot\theta^2
\\
L_B[\theta,\mathbf R(\theta)]&=\sum_{j=1}^\mathcal N\frac{m}{2}\dot R_j(\theta)^2-\sum_{j=1}^\mathcal N\frac{1}{2}m\omega_j^2\left(R_j(\theta)-\frac{C_j\theta}{m\omega_j^2}\right)^2
\end{align}
is the Lagrangian of the coupled system.
The action $S[\theta,\mathbf R(\theta)]$ is given by
\begin{align}
S[\theta,\mathbf R(\theta)]=\int_0^tL[\theta,\mathbf R(\theta)]ds&=S_0[\theta]+S_B[\theta,\mathbf R(\theta)]
\\
S_0[\theta]&=\int_0^td\tau L[\theta]
\\
S_B[\theta,\mathbf R(\theta)]&=\int_0^td\tau L_B[\theta,\mathbf R(\theta)].
\label{S_0}
\end{align}
Now, suppose that we initially have the total density operator given by
\begin{equation}
\hat\rho_{AB}(0)=\hat\rho_A(0)\hat\rho_B(0).
\end{equation}
Then, the reduced density matrix is
\begin{align}
\rho(\theta,\phi,t)&=\int_{-\infty}^\infty d\mathbf R(\theta)d\mathbf Q(\theta)\delta(\mathbf R(\theta)-\mathbf Q(\theta))\rho(\theta,\mathbf p,\phi,\mathbf s,t)
\nonumber\\
&=\int_0^{2\pi}d\theta' d\phi' J(\theta,\phi,t;\theta',\phi',0)\rho_A(\theta',\phi',0)
\label{Reduced_Density_Matrix}
\end{align}
where
\begin{equation}
J(\theta,\phi,t;\theta',\phi',0)=\sum_{l=-\infty}^\infty\sum_{l'=-\infty}^\infty\int_{\theta'}^{\theta+2\pi l}\mathcal D\theta\int_{\phi'}^{\phi+2\pi l'}\mathcal D\phi^\ast\exp\frac{i}{\hbar}\left(S_0[\theta]-S_0[\phi]\right)\mathcal F[\theta,\phi]
\end{equation}
is the propagator of the density matrix,
\begin{align*}
\mathcal F[\theta,\phi]&=\int_{-\infty}^\infty d\mathbf R(\theta)d\mathbf Q(\phi) d\mathbf R'(\theta') d\mathbf Q'(\phi')\delta(\mathbf R(\theta)-\mathbf Q(\phi))\rho_B(\mathbf p',\mathbf s',0)
\\
&\times \int_{\mathbf R'(\theta')}^{\mathbf R(\theta)}\mathcal D\mathbf R(\theta)\int_{\mathbf Q'(\phi')}^{\mathbf Q(\phi)}\mathcal D\mathbf Q(\phi)\exp\frac{i}{\hbar}\left(S_B[\theta,\mathbf R(\theta)]-S_B[\phi,\mathbf Q(\phi)]\right)
\end{align*}
is the influence functional and $Q(\phi)=-(s_j-c\phi)/m\omega_j$. 
Supposed that the density operator $\hat \rho_B$ can be written as a canonical ensemble at $t=0$. That is,
\begin{equation}
\rho_B=\frac{\exp(-\beta\hat H_B)}{\mathrm{tr}_B[\exp(-\beta\hat H_B)]}
\end{equation}
Introduce the imaginary time $\tau=-i\hbar\beta$, then the density matrix of the environment at $t=0$ is
\begin{align*}
\rho_B(\mathbf p',\mathbf s',0)=\frac{\Braket{\mathbf p'|\exp\left(-\frac{i\tau}{\hbar}\hat H_B\right)|\mathbf s'}}{\mathrm{tr}_B\left[\exp\left(-\frac{i\tau}{\hbar}\hat H_B\right)\right]}=\frac{K_B(\mathbf p',\tau;\mathbf s',0)}{\int d\mathbf qK_B(\mathbf q,\tau;\mathbf q,0)}.
\end{align*}

\label{F_Derivation}
\subsection{Momentum Representation of the Density matrix of a Harmonic Oscillator}

The propagator of a harmonic oscillator with  the Lagrangian
\begin{equation}
L=\frac{m}{2}\dot x^2-\frac{1}{2}m\omega x^2
\end{equation}
has a well known solution
\begin{align}
K(x,t;x',0)&=\int \mathcal Dx\exp\frac{i}{\hbar}\int_0^tLds
\\
&=\sqrt{\frac{m\omega}{2\pi i\hbar\sin\omega t}}\exp\left\{\frac{i}{\hbar}\frac{m\omega}{2\sin\omega t}[(x^2+x'^2)\cos\omega t-2xx']\right\}
\end{align}
%
$K(p,t;p',0)$ is obtained by a Fourier transformation
\begin{equation}
K(p,t:p',0)=\int_{-\infty}^\infty\frac{dxdx'}{2\pi\hbar}K(x,t;x',0)\exp\frac{i}{\hbar}\left(px-p'x'\right).
\end{equation}
Let $A=\frac{m\omega\cos\omega t}{2\sin\omega t}$ and $B=\frac{m\omega}{2\sin\omega t}$, then
\begin{align*}
K(p,t;p',0)=\frac{F(t)}{\sqrt{4B^2-4A^2}}\exp\frac{i}{\hbar}\frac{Ap^2+Ap'^2-2Bpp'}{4B^2-4A^2}
\end{align*}
where
\begin{equation}
4B^2-4A^2=\frac{m^2\omega^2(1-\cos^2\omega t)}{\sin^2\omega t}=m^2\omega^2,
\end{equation}
so,
\begin{equation}
K(p,t;p',0)=\frac{F(t)}{m\omega}\exp\frac{i}{\hbar}S[p(t)/m\omega].
\end{equation}
Therefore, using $\tau=-i\hbar\beta$, $\beta=1/k_BT$, $\nu_j=\hbar\omega_j\beta$ and $i\sinh x=\sin ix$ we have
\begin{align}
\braket{\mathbf p'|\exp(-i\tau\hat H_B/\hbar)|\mathbf s'}=\prod_{j=1}^\mathcal{N}\frac{1}{m\omega_j}\sqrt{\frac{m\omega_j}{2\pi \hbar\sinh\nu_j}}\exp-\frac{m\omega_j}{2\hbar\sinh\nu_j}\left[\frac{{p'}_j^2+{s'}_j^2}{m^2\omega_j^2}\cosh\nu-2\frac{{p'}_j{s'}_j}{m^2\omega_j^2}\right],
\end{align}
and
\begin{align*}
\mathrm{tr}_B[\exp(-i\tau\hat H_B/\hbar)]=\int_{-\infty}^\infty d\mathbf q\braket{\mathbf q|\exp(-i\tau\hat H_B/\hbar)|\mathbf q}=\prod_{j=1}^\mathcal{N}\frac{1}{2\sinh\frac{\nu_j}{2}}.
\end{align*}
We are assuming that the environment is not coupled to the system at $t=0$, so $R'_j=p'_j/m\omega_j$ and $Q'_j=s'_j/m\omega_j$. Therefore,
\begin{equation}
\rho_B(\mathbf p',\mathbf s',0)=\left(\prod_{j=1}^\mathcal{N}\frac{1}{m\omega_j}\right)\rho_B(\mathbf R',\mathbf Q',0)
\end{equation}
where
\begin{equation}
\rho_B(\mathbf R',\mathbf Q',0)=\prod_{j=1}^\mathcal{N}2\sinh\frac{\nu_j}{2}\sqrt{\frac{m\omega_j}{2\pi \hbar\sinh\nu_j}}\exp-\frac{m\omega_j}{2\hbar\sinh\nu_j}\left[({R'}_j^2+{Q'}_j^2)\cosh\nu_j-2{R'}_j{Q'}_j\right].
\end{equation}
Let us rewrite this density matrix into a simpler form. Define $x_{0j}=R'_j+Q'_j$ and $x'_{0j}=R'_j-Q'_j$. Then, using
\begin{equation}
({R'_j}^2+{Q'_j}^2)\cosh\nu_j-2{R'}_j{Q'}_j=\frac12(\cosh\nu_j-1)x_{0j}^2+\frac12(\cosh\nu_j+1){x'_{0j}}^2
\end{equation}
and the identity
\begin{equation}
\frac{\cosh\nu_j-1}{\sinh\nu_j}=\frac{\sinh\nu_j}{\cosh\nu_j+1}=\tanh\frac{\nu_j}{2},
\end{equation}
we obtain
\begin{equation}
\rho_B(\mathbf R',\mathbf Q',0)=\prod_{j=1}^\mathcal{N}\sqrt{\frac{m\omega_j}{\pi \hbar\mu_j}}\exp-\frac{m\omega_j}{4\hbar}\left[\frac{x_{0j}^2}{\mu_j}+{x'_{0j}}^2\mu_j\right]. \label{densitmatrix RQ}
\end{equation}
\begin{equation}
\mu_j=\coth\frac{\nu_j}{2}.
\end{equation}
\subsection{The Influence Functional $\mathcal F[\theta,\phi]$}
Now, we calculate the density functional $\mathcal F[\theta,\phi]$ explicitly. We will write $\mathbf R$ and $\mathbf Q$ instead of $\mathbf R[\theta]$ and $\mathbf Q[\phi]$ for simplicity, but keep in mind that they depend on $\theta$ and $\phi$, respectively. Then, the influence functional can be written as
\begin{align}
\mathcal F[\theta,\phi]&=\int_{-\infty}^\infty d\mathbf x_t d\mathbf x'_td\mathbf x_0d\mathbf x'_0(1/2)^{\mathcal{N}}\delta(\mathbf x'_t)\rho_B(\mathbf R',\mathbf Q',0)\int_{\mathbf R'}^{\mathbf R}\mathcal D\mathbf R\int_{\mathbf Q'}^{\mathbf Q}\mathcal D\mathbf Q^\ast\nonumber\\
&\times\exp\left\{\frac{i}{\hbar}\int_0^tds\sum_j\left[\frac{m}{2}(\dot R_j^2-\dot Q_j^2)-\frac{1}{2}m\omega_j^2(R_j^2-Q_j^2)+C_j(\theta R_j-\phi Q_j)-\frac{C_j^2}{2m\omega_j^2}(\theta^2-\phi^2)\right]\right\}.
\end{align}
To evaluate this we introduce the variables
\begin{align}
\begin{split}
\varphi&=\theta+\phi,\qquad \varphi'=\theta-\phi,
\\
x_j&=R_j+Q_j,\qquad x'_j=R_j-Q_j.
\end{split}
\end{align}
Then, the influence functional becomes
\begin{align*}
\mathcal F[\theta,\phi]&=\int_{-\infty}^\infty d\mathbf x_td\mathbf x'_td\mathbf x_0d\mathbf x'_0(1/2)^{\mathcal{N}}\delta(\mathbf x'_t)\rho_B(\mathbf R',\mathbf Q',0)\int\mathcal D\mathbf x\mathcal D \mathbf x'
\\
&\times\exp\left\{\frac{i}{\hbar}\int_0^tds\sum_j\left[\frac{m}{2}\dot x_j\dot x'_j-\frac{1}{2}m\omega_j^2x_jx'_j+\frac{C_j}{2}(\varphi x'_j+\varphi'x_j)-\frac{C_j^2}{2m\omega_j^2}\varphi'\varphi\right]\right\}
\\
&=\int_{-\infty}^\infty d\mathbf x_td\mathbf x'_td\mathbf x_0d\mathbf x'_0(1/2)^{\mathcal{N}}\delta(\mathbf x'_t)\rho_B(\mathbf R',\mathbf Q',0)
\\
&\times\exp\left\{\frac{im}{2\hbar}\sum_j(x_{tj}\dot x'_{tj}-x_{0j}\dot x'_{0j})-\frac{i}{\hbar}\int_0^tds\sum_j\frac{C_j^2}{2m\omega_j^2}\varphi'\varphi\right\}
\\
&\times\int\mathcal D\mathbf x\mathcal D \mathbf x'\exp\left\{\frac{i}{\hbar}\int_0^tds\sum_j\left[-g(x'_j)x_j+\frac{C_j}{2}\varphi x_j'\right]\right\}
\end{align*}
where
\begin{equation}
g(x_j')=\frac{m}{2}\ddot x'_j+\frac{1}{2}m\omega_j^2x'_j-\frac{C_j}{2}\varphi'.
\end{equation}
We note that the classical solution $x_{j,cl}'$ satisfies the Euler-Lagrange equation
\begin{equation}
g(x_{j,cl}')=0,\quad g(x_{j}'(0))=0,\quad g(x_{j}'(t))=0.
\end{equation}
The path integral over $\mathbf x$ can be done first.  Calling this integral $I_{\mathbf x,\mathbf x'}$ we obtain
\begin{align}
I_{\mathbf x,\mathbf x'}&=\int\mathcal D\mathbf x\mathcal D \mathbf x'\exp\left\{\frac{i}{\hbar}\int_0^tds\sum_j\left[-g(x'_{j})x_{j}+\frac{C_j}{2}\varphi x_{j}'\right]\right\}
\\
&=\frac{1}{|\mathcal A_B|^2}\int_{-\infty}^\infty\left(\prod_{n=1}^{N-1}\frac{d\mathbf xd\mathbf x'}{|\mathcal A_B|^2}\right)\prod_{n=1}^N\prod_j\exp\frac{i\epsilon}{\hbar}\left[-g(x_{n,j}')x_{n,j}+\frac{C_j}{2}\varphi x_j'\right]
\\
&=\frac{1}{|\mathcal A_B|^2}\int_{-\infty}^\infty\left(\prod_{n=1}^{N-1}\frac{d\mathbf x'}{|\mathcal A_B|^2}\right)
\prod_j\left(\prod_{n=1}^{N-1}\delta\left[\frac{\epsilon}{2\pi\hbar}g(x'_{n,j})\right]\right)\prod_{n=1}^N\exp\frac{i\epsilon}{\hbar}\frac{C_j}{2}\varphi x_{n,j}'.
\end{align}
$|\mathcal A|^{-2}$ is absorbed into the delta functions. The interpretation of the delta function is to insert the classical solution $x'_{j,cl}$ into $x'_j$ which satisfy $g(x'_{j,cl})=0$. The outcome is, 
\begin{equation}
I_{\mathbf x,\mathbf x'}=\frac{1}{|\mathcal A_B|^2}\exp\left(\frac{i}{\hbar}\int_0^tds\sum_j\frac{C_j}{2}\varphi x'_{cl}
\right).
\end{equation}
Therefore, the influence functional is
\begin{align*}
\mathcal F[\theta,\phi]&=\int_{-\infty}^\infty d\mathbf x_td\mathbf x'_td\mathbf x_0d\mathbf x'_0(1/2)^{\mathcal{N}}\delta(\mathbf x'_t)\rho_B(\mathbf R',\mathbf Q',0)\exp\frac{i}{\hbar}\frac{m}{2}\sum_j(x_{tj}\dot x'_{tj}-x_{0j}\dot x'_{0j})
\\
&\times\frac{1}{|\mathcal A_B|^2}\exp\left(\frac{i}{\hbar}\int_0^tds\sum_j\left[-\frac{C_j^2}{2m\omega_j^2}\varphi'\varphi+\frac{C_j}{2}\varphi x_{cl}'\right]\right).
\end{align*}
The $\mathbf x_t$ integral gives a delta function of $\dot{\mathbf x}'_t$
\begin{align*}
\mathcal F[\theta,\phi]&=\int_{-\infty}^\infty d\mathbf x'_td\mathbf x_0d\mathbf x'_0(1/2)^{\mathcal{N}}\delta(\mathbf x'_t)\delta(\dot{\mathbf x}'_t)\rho_B(\mathbf R',\mathbf Q',0)
\\
&\times\exp\left(\frac{i}{\hbar}\int_0^tds\sum_j\left[-\frac{C_j^2}{2m\omega_j^2}\varphi'\varphi+\frac{C_j}{2}\varphi x_{cl}'\right]+\frac{i}{\hbar}\frac{m}{2}\sum_jx_{0j}\dot x'_{0j}\right).
\end{align*}
Insert equation \eqref{densitmatrix RQ} for the density matrix and do the $\mathbf x_0$ integral, then we obtain
\begin{align*}
\mathcal F[\theta,\phi]&=\int_{-\infty}^\infty d\mathbf x'_td\mathbf x'_0\delta(\mathbf x'_t)\delta(\dot{\mathbf x}'_t)
\\
&\times\prod_j\exp\left(\frac{i}{\hbar}\int_0^tds\left[-\frac{C_j^2}{2m\omega_j^2}\varphi'\varphi+\frac{C_j}{2}\varphi x_{cl}'\right]-\frac{m\mu_j\omega_j}{4\hbar}\left(\frac{\dot{x}{'_{0j}}^2}{\omega_j^2}+{x'_{0j}}^2\right)\right).
\end{align*}
The two delta functions $\delta(\mathbf x'_t)$ and $\delta(\dot{\mathbf x}'_t)$ can be interpreted as boundary conditions on the classical solution of $g(\mathbf x'(s))=0$. The result of doing the remaining integrals would be to substitute the classical solution of $x'_{cl}$, $\dot{x}{'_{0j}}$ and $x'_{0j}$ which satisfy these boundary conditions. 
The solution to the classical solution of $g(x)=0$, that is
\begin{equation}
\ddot x'_j+\omega_j^2x'_j-\frac{C_j}{m}\varphi'=0.
\end{equation}
is
\begin{align}
x_j(\tau)&=-\int_\tau^t\frac{C_j^2\varphi'(s)}{m\omega_j}\sin\omega_j(\tau-s)ds+x'_{tj}\cos\omega_j(t-\tau)-\frac{\dot x'_{tj}}{\omega_j}\sin\omega_j(t-\tau).
\end{align}
For our boundary condition this solution reduces to
\begin{align}
x'_j(\tau)&=-\int_\tau^t\frac{C_j\varphi'(s)}{m\omega_j}\sin\omega_j(\tau-s)ds
\\
\dot{x}'_j(\tau)&=-\int_\tau^t\frac{C_j\varphi'(s)}{m}\cos\omega_j(\tau-s)ds
\end{align}
Therefore, the result of integration is
\begin{align*}
\mathcal F[\theta,\phi]=\exp\left(-\sum_j\frac{i}{\hbar}\frac{C_j^2}{2m\omega_j}\int_0^t\int_\tau^t\varphi(\tau)\left(\frac{2}{\omega_j}\delta(\tau-s)-\sin\omega_j(\tau-s)\right)\varphi'(s)d\tau ds\right)
\\
\times\exp\left(-\sum_j\frac{\mu_j}{4\hbar}\int_0^t\int_0^t\frac{C_j^2}{m\omega_j}\varphi'(\tau)\cos\omega_j(\tau-s)\varphi'(s)dsd\tau\right).
\end{align*}
Since $s$ and $\tau$ enter into the second integral symmetrically it can be rewritten
\begin{equation}
\int_0^t\int_0^tE(\tau,s)dsd\tau=2\int_0^t\int_0^\tau E(\tau,s)dsd\tau.
\end{equation}
The first integral can be written
\begin{equation}
\int_0^t\int_\tau^tE(\tau,s)dsd\tau=\int_0^t\int_0^\tau E(s,\tau)dsd\tau.
\end{equation}
Then, the influence functional becomes
\begin{equation}
\mathcal F[\theta,\phi]=\exp i\Phi[\theta,\phi]
\end{equation}
where the influence phase $\Phi[\theta,\phi]$ can be written as
\begin{align}
i\Phi[\theta,\phi]&=-\frac{i}{\hbar}\int_0^t\int_0^\tau d\tau ds\varphi(s)\alpha_I(\tau-s)\varphi'(\tau)-\frac{1}{\hbar}\int_0^t\int_0^\tau dsd\tau\varphi'(\tau)\alpha_R(\tau-s)\varphi'(s)
\\
\alpha_R(\tau-s)&=\sum_j\frac{C_j^2}{2m\omega_j}\coth\frac{\hbar\omega_j}{2k_BT}\cos\omega_j(\tau-s)
\\
\alpha_I(\tau-s)&=\sum_j\frac{C_j^2}{2m\omega_j}\left(\frac{2}{\omega_j}\delta(\tau-s)+\sin\omega_j(\tau-s)\right).
\end{align}
Therefore, we obtain the density matrix propagator $J(\theta,\phi,t;\theta',\phi',0)$
\begin{align}
&J(\theta_\text f,\phi_\text f,t;\theta\text i,\phi_\text i,0)
\\
&=\left.
\int_{\theta_\text i}^{\theta_\text f}D\theta\int_{\phi_\text i}^{\phi_\text f}D\phi^\ast\exp\left\{\frac{i}{\hbar}S[\varphi^+,\varphi^-]-\Gamma[\varphi^-]\right\}
\right|_{\substack{\varphi^+=(\theta+\phi)/2\\\varphi^-=\phi-\theta\quad}},
\label{JDef}
\end{align}
the classical action $S[\varphi^+,\varphi^-]$ is
\begin{align*}
S[\varphi,\varphi']&=-\int_0^td\tau I\dot\varphi^+(\tau)\dot\varphi^-(\tau)
\\
&+2\int_0^td\tau\int_0^s ds\varphi^-(\tau)\alpha_\text I(\tau-s)\varphi^+(s)
\end{align*}
and the $\Gamma[\varphi^-]$ is given by 
\begin{align*}
\Gamma[\varphi^-]&=\frac{1}{2\hbar}\int_0^td\tau\int_0^tds\varphi^-(\tau)\alpha_\text R(\tau-s)\varphi^-(s).
\end{align*}
The path integral in \eqref{JDef} can be done exactly and we obtain
\begin{align}
J(\theta_\text f,\phi_\text f,t;\theta_\text i,\phi_\text i,0)&=F^2(t)\exp\left(\frac{i}{\hbar}S[\varphi^+_\text{cl},\varphi^-_\text{cl}]-\Gamma[\varphi^-_\text{cl}]\right)\label{JClassical}
\end{align}
$\varphi^\pm_\text{cl}$ are the classical coordinates obtained from the Euler-Lagrange equation
\begin{align}
I\ddot\varphi_\text{cl}^-(u)+2\int_u^td\tau\varphi_\text{cl}^-(\tau)\alpha_\text I(\tau-u)=0\label{eqn1}
\\
I\ddot\varphi^+_\text{cl}(u)+2\int_0^ud\tau\varphi^+_\text{cl}(\tau)\alpha_\text I(u-\tau)=0\label{eqn2}
\end{align}
whose solution are given in terms of boundary conditions $\varphi^\pm_\text i=\varphi^\pm(0),\varphi^\pm_\text f=\varphi^\pm(t)$:
\begin{align}
\varphi^+_\text{cl}(u)&=\kappa_i(u;t)\varphi^+_\text i+\kappa_f(u;t)\varphi^+_\text f
\label{varphisolfinal}
\\
\varphi^-_\text{cl}(u)&=\kappa_i(t-u;t)\varphi^-_\text f+\kappa_f(t-u;t)\varphi^-_\text i
\label{varphiprimesolfinal}
\\
\kappa_i(u;t)&=\dot G(u)-\frac{\dot G(t)}{G(t)}G(u),
\qquad
\kappa_f(u;t)=\frac{G(u)}{G(t)}.
\end{align}
Then, the classical action reduces to
\begin{align}
S[\varphi^+_\text{cl},\varphi^-_\text{cl}]&=-I[\dot\varphi^+_\text{cl}(t)\varphi^-_\text f-\dot\varphi^-_\text{cl}(0)\varphi^-_\text i].
\label{ClassicalAction}
\end{align}
\end{document}